\documentstyle[12pt]{article}

\def\hybrid{\topmargin -20pt	\oddsidemargin 0pt
	\headheight 0pt	\headsep 0pt
	\textwidth 6.25in	
	\textheight 9.5in	
	\marginparwidth .875in
	\parskip 5pt plus 1pt	\jot = 1.5ex}
\def\cQ{{\cal Q}}
\def\cG{{\cal G}}
\def\cL{{\cal L}}
\def\cH{{\cal H}}
\def\ket#1{|{#1}\rangle}
\def\noi{\noindent}
\def\half{{1\over2}}
\def\baselinestretch{1.2}

\catcode`\@=11

\def\marginnote#1{}
\def\draftlabel#1{{\@bsphack\if@filesw {\let\thepage\relax
   \xdef\@gtempa{\write\@auxout{\string
      \newlabel{#1}{{\@currentlabel}{\thepage}}}}}\@gtempa
   \if@nobreak \ifvmode\nobreak\fi\fi\fi\@esphack}
	\gdef\@eqnlabel{#1}}
\def\@eqnlabel{}
\def\@vacuum{}
\def\draftmarginnote#1{\marginpar{\raggedright\scriptsize\tt#1}}

\def\draft{\oddsidemargin -.2truein
	\def\@oddfoot{\sl preliminary draft \hfil
	\rm\thepage\hfil\sl\today\quad\militarytime}
	\let\@evenfoot\@oddfoot	\overfullrule 3pt
	\let\label=\draftlabel
	\let\marginnote=\draftmarginnote
   \def\@eqnnum{(\theequation)\rlap{\kern\marginparsep\tt\@eqnlabel}%
\global\let\@eqnlabel\@vacuum}  }

\def\preprint{\twocolumn\sloppy\flushbottom\parindent 2em
	\leftmargini 2em\leftmarginv .5em\leftmarginvi .5em
	\oddsidemargin -.5in	\evensidemargin -.5in
	\columnsep .4in	\footheight 0pt
	\textwidth 10.in	\topmargin  -.4in
	\headheight 12pt \topskip .4in
88	\textheight 6.9in \footskip 0pt
	\def\@oddhead{\thepage\hfil\addtocounter{page}{1}\thepage}
	\let\@evenhead\@oddhead	\def\@oddfoot{}	\def\@evenfoot{} }

\def\numberbysection{\@addtoreset{equation}{section}
	\def\theequation{\thesection.\arabic{equation}}}

\def\underline#1{\relax\ifmmode\@@underline#1\else
	$\@@underline{\hbox{#1}}$\relax\fi}

\def\titlepage{\@restonecolfalse\if@twocolumn\@restonecoltrue
\onecolumn
     \else \newpage \fi \thispagestyle{empty}\c@page\z@
	\def\thefootnote{\fnsymbol{footnote}} }

\def\endtitlepage{\if@restonecol\twocolumn \else \newpage \fi
	\def\thefootnote{\arabic{footnote}}
	\setcounter{footnote}{0}}  

\catcode`@=12
\relax

\def\figcap{\section*{Figure Captions\markboth
	{FIGURECAPTIONS}{FIGURECAPTIONS}}\list
	{Figure \arabic{enumi}:\hfill}{\settowidth\labelwidth{Figure
999:}
	\leftmargin\labelwidth
	\advance\leftmargin\labelsep\usecounter{enumi}}}
 \relax
\def\tablecap{\section*{Table Captions\markboth
	{TABLECAPTIONS}{TABLECAPTIONS}}\list
	{Table \arabic{enumi}:\hfill}{\settowidth\labelwidth{Table
999:}
	\leftmargin\labelwidth
	\advance\leftmargin\labelsep\usecounter{enumi}}}
 \relax
\def\reflist{\section*{References\markboth
	{REFLIST}{REFLIST}}\list
	{[\arabic{enumi}]\hfill}{\settowidth\labelwidth{[999]}
	\leftmargin\labelwidth
	\advance\leftmargin\labelsep\usecounter{enumi}}}
 \relax
\makeatletter
\newcounter{pubctr}
\def\publist{\@ifnextchar[{\@publist}{\@@publist}}
\def\@publist[#1]{\list
	{[\arabic{pubctr}]\hfill}{\settowidth\labelwidth{[999]}
	\leftmargin\labelwidth
	\advance\leftmargin\labelsep
	\@nmbrlisttrue\def\@listctr{pubctr}
	\setcounter{pubctr}{#1}\addtocounter{pubctr}{-1}}}
\def\@@publist{\list
	{[\arabic{pubctr}]\hfill}{\settowidth\labelwidth{[999]}
	\leftmargin\labelwidth
	\advance\leftmargin\labelsep
	\@nmbrlisttrue\def\@listctr{pubctr}}}
 \relax
\makeatother
\newskip\humongous \humongous=0pt plus 1000pt minus 1000pt

\newif\ifdtup

\font\Scbig=cmss10 scaled\magstep1
\font\Scscr=cmss8 scaled\magstep1
\font\Scscrscr=cmss8
\newfam\Scfam
\textfont\Scfam=\Scbig
\scriptfont\Scfam=\Scscr
\scriptscriptfont\Scfam=\Scscrscr
\def\Sc{\fam\Scfam}

\relax
\hybrid
\def\lvm{\leavevmode\hbox to\parindent{\hfill}}

\def\thefootnote{\fnsymbol{footnote}}
\def\BE{\begin{equation}}
\def\EE{\end{equation}}
\def\BA{\begin{eqnarray}}
\def\EA{\end{eqnarray}}

\def\G{\Gamma}

\def\a{\alpha}
\def\th{\theta}

\def\P{\Phi}

\def\tt{\bar\tau}

\def\lvm{\leavevmode\hbox to\parindent{\hfill}}
\def\bar{\overline}
\def\req#1{(\ref{#1})}

\def\L{\left}
\def\R{\right}

\def\BE{\begin{equation}}
\def\EE{\end{equation} \vskip 0.30\baselineskip}
\def\BA{\begin{array}}
\def\EA{\end{array}}

\def\noi{\noindent}

\def\frac#1#2{{\textstyle{{#1}\over{#2}}}}
\def\half{{1\over2}}

\def\Kr#1{\delta_{{#1},0}}
\def\ket#1{|{#1}\rangle}

\def\cG{{\cal G}}
\def\cH{{\cal H}}

\def\cL{{\cal L}}

\def\cQ{{\cal Q}}

\def\open#1{\mbox{{\bf{#1}}}}

\def\oZ{{\open Z}}

\def\ctop{{\Sc c}}
\def\htop{{\Sc h}}

\def\a{\alpha}
\def\b{\beta}
\def\g{\gamma}
\def\Ups{\Upsilon}

\newif\ifold \oldtrue \def\new{\oldfalse}
\let\ssection=\section
\def\section{\setcounter{equation}{0}\ssection}

\begin{document}
\renewcommand{\theequation}{\thesection.\arabic{equation}}
\newcommand{\beq}{\begin{equation}}
\newcommand{\eeq}[1]{\label{#1}\end{equation}}
\newcommand{\ber}{\begin{eqnarray}}
\newcommand{\eer}[1]{\label{#1}\end{eqnarray}}
\begin{titlepage}
\begin{center}

\hfill IMAFF-94/5\\
\hfill hep-th/9411119\\
\vskip .5in

{\large \bf Topological Descendants: DDK and KM Realizations}
\vskip .8in

{\bf Beatriz Gato-Rivera and Jose Ignacio Rosado}\\
\vskip
 .3in

{\em Instituto de Matem\'aticas y F\'\i sica Fundamental, CSIC,\\ Serrano 123,
Madrid 28006, Spain} \footnote{e-mail addresses:
bgato, jirs @cc.csic.es}\\

\vskip .5in

\end{center}

\vskip .6in

\begin{center} {\bf ABSTRACT } \end{center}
\begin{quotation}
The "minimal matter + scalar" system can be embedded into the twisted
$N=2$ topological algebra in two ways: \`a la DDK or \`a la KM.
Here we present some results concerning the topological descendants
and their DDK and KM realizations.
In particular, we prove four "no-ghost" theorems (two for
null states) regarding the
reduction of the topological descendants into secondaries of
the "minimal matter + scalar" conformal field theory. We
 write down  the relevant
expressions for the case of level $2$ descendants.
\end{quotation}
\vskip 1.5cm

November 1994\\
\end{titlepage}
\vfill
\eject
\def\baselinestretch{1.2}
\baselineskip 16 pt
\section{Introduction}\lvm

Two years ago it was shown \cite{BeSe2}, \cite{BeSe3} that the
"$d \leq 1$ matter + scalar" system, extended by appropriate
$bc$ ghosts, provides two different realizations of the twisted
$N=2$ topological algebra: the DDK (David-Distler-Kawai) and
the KM (Kontsevich-Miwa) realizations. The first one was a
bosonic string construction with all the elements therein
(the Liouville field and the $c=-26$ reparametrization ghosts)
\cite{DDK},
while the second was related to the KP hierarchy
through the Kontsevich-Miwa transform  \cite{BeSe3},\cite{SeKM}
 (for a short, pedagogical
introduction to the subject see \cite{BJI1}).

In this letter we address the issue of the topological descendants
(bosonic as well as fermionic)
and their special features when they are realized  \`a la DDK or
\`a la KM. In particular, we prove four
{\it no-ghost} theorems concerning the
 reduction of the topological descendants into secondaries of
the "matter + scalar" conformal field theory (two of the theorems
deal with null descendants).
 A key fact proves to be the $\cQ_{n \geq 0}$  or $\cG_{n \geq 0}$
invariance of the topological states ($\cQ_0$ or $\cG_0$ invariance
of the topological null states).
 As an example, we write down
the relevant results for the case of level $2$ descendants,
showing complete agreement with the {\it no-ghost} theorems.

\section{Topological Descendants}\lvm

The $N=2$ twisted topological algebra reads \cite{[EY]},
\cite{[W-top]}
\BE\new\BA{lclclcl}
\L[\cL_m,\cL_n\R]&=&(m-n)\cL_{m+n}\,,&\qquad&[\cH_m,\cH_n]&=
&{\ctop\over3}m\Kr{m+n}\,,\\
\L[\cL_m,\cG_n\R]&=&(m-n)\cG_{m+n}\,,&\qquad&[\cH_m,\cG_n]&=&\cG_{m+n}\,,
\\
\L[\cL_m,\cQ_n\R]&=&-n\cQ_{m+n}\,,&\qquad&[\cH_m,\cQ_n]&=&-\cQ_{m+n}\,,\\
\L[\cL_m,\cH_n\R]&=&\multicolumn{5}{l}{-n\cH_{m+n}+{\ctop\over6}(m^2+m)
\Kr{m+n}\,,}\\
\L\{\cG_m,\cQ_n\R\}&=&\multicolumn{5}{l}{2\cL_{m+n}-2n\cH_{m+n}+
{\ctop\over3}(m^2+m)\Kr{m+n}\,,}\EA\qquad m,~n\in\oZ\,.\label{topalgebra}
\EE

\noi
where $\cL_m$ and $\cH_m$ are the bosonic generators corresponding
to the energy momentum tensor (Virasoro generators)
 and the topological $U(1)$ current respectively, while
$\cQ_m$ and $\cG_m$ are the fermionic generators corresponding
to the BRST current and the spin-2 fermionic current
 respectively. The "topological central
charge" $\ctop$ is the true central charge of the $N=2$
superconformal algebra \cite{LVW} .

In what follows we will restrict ourselves to descendants built on
 chiral primary states, that is, highest weight states that are
also $\cQ_0$ and $\cG_0$ invariant.

\newpage

\subsection{Some Remarks}\lvm

Let us consider the anticommutator
$\{\cQ_0, \cG_0\} = 2 \cL_0$ between the BRST charge $\cQ_0$ and
$\cG_0$. Acting on chiral primary states
 this relation implies zero conformal dimension
for the corresponding fields. Acting on secondary
 states, on the other hand,
this anticommutator gives very useful
information:

 First, it shows that a $\cQ_0$-closed ($\cG_0$-closed)
secondary state is also $\cQ_0$-exact ($\cG_0$-exact)

\BE\ket\Xi = {1\over{2\Delta}} (\cQ_0 \cG_0 + \cG_0 \cQ_0)
 \ket\Xi \label{QG} \EE

\noi
($\Delta$ is the conformal dimension of the corresponding secondary field).

Secondly, it tells us that bosonic and fermionic descendants are
related in a peculiar way. Namely, $\cQ_0$-invariant bosonic
(fermionic) states are $\cG_0$ mapped onto
 $\cG_0$-invariant fermionic (bosonic)
states that in turn are $\cQ_0$ mapped onto the original
 $\cQ_0$-invariant bosonic (fermionic) ones (up to a scale factor
of $2 \Delta$). Similarly,
$\cG_0$-invariant bosonic (fermionic) states are $\cQ_0$ mapped onto
$\cQ_0$-invariant fermionic (bosonic) that in turn are $\cG_0$
mapped onto the original $\cG_0$-invariant bosonic (fermionic) states.

In particular, bosonic and fermionic topological null vectors
are connected in this way, since the $\cG_0$ and $\cQ_0$ actions
on  null vectors give  null vectors as well.

\subsection{DDK and KM Realizations of Topological Descendants}\lvm

The DDK and KM realizations of the topological algebra
\req{topalgebra} are the two possible twistings of the
same $N=2$ superconformal theory \cite{BeSe3}.
The field content in both realizations is very similar:
$d \leq 1$ matter +  scalar +  $bc$ system.
However, the scalars differ by the background charge and the
 way they dress the matter $(Q_s = Q_{Liouville},\  \Delta = 1$ in
the DDK case versus $Q_s = Q_{matter},\  \Delta = 0 $ in the KM
case), while the $bc$ systems differ by the spin and the central
charge $(s=2,\  c=-26$ for the DDK ghosts
 versus $s=1,\  c=-2$ for the KM ghosts).

\vskip .2in

In the DDK realization the generators of the topological algebra are

\BE \cL_m=L_m+l_m,\quad
l_m\equiv\sum_{n\in{\oZ}}(m+n):\!b_{m-n}c_n\!:\label{L26}\EE

\BE\cH_m=\sum_{n\in\oZ}:\!b_{m-n}c_n\!:{}-
{}\sqrt{{3-\ctop\over3}}I_m
\,,\label{H26}\EE

\BE\cQ_m=2\sum_{p\in\oZ}c_{m-p}L_p
+\sum_{p,r\in\oZ}(p-r):\!b_{m-p-r}c_pc_r\!:{}+
{}2\sqrt{{3-\ctop\over3}}m
\sum_{p\in\oZ}c_{m-p}I_p+{\ctop\over3}(m^2-m)c_m~,\label{Q26}\EE

\BE\cG_m=b_m\,,\label{G26}\EE

\noi
and the chiral primary states can be written as
$\ket\P = \ket\Ups \otimes c_1 \ket0_{gh}$,
where $\ket\Ups$ is a primary state in the "matter + scalar" sector
(for a spin-2 $bc$ system $c_1 \ket0_{gh}$ is the "true" ghost vacuum
annihilated by all the positive modes $b_n$ and $c_n$).

In the KM realization the generators read

\BE\cL_m=L_m+l_m,\quad
l_m=\sum_{n\in{\oZ}}n\!:\!b_{m-n}c_n\!: \label{L}\EE

\BE\cH_m= - \sum_{n\in\oZ}:\!b_{m-n}c_n\!: +
{}\sqrt{{3-\ctop\over3}}I_m
\,,\label{H}\EE

\BE\cQ_m=b_m\,,\label{Q}\EE

\BE\cG_m=2\sum_{p\in\oZ}c_{m-p}L_p+2{}\sqrt{{3-\ctop\over3}}
\sum_{p\in\oZ}(m-p)c_{m-p}I_p
{}+\sum_{p,r\in\oZ}(r-p):\!b_{m-p-r}c_rc_{p}\!:
{}+{}{\ctop\over3}(m^2+m)c_m\,,\label{G}\EE

\noi
and the chiral primary states split as $\ket\P = \ket\Ups \otimes
\ket0_{gh}$.

\vskip .2in

Now let us define the DDK and KM conformal field theories (CFT's) as
the theories given by the "matter + scalar" systems, without
the ghosts. These theories are described by the commutation
relations

 \BE\new\BA{rcl}
\L[L_m,L_n\R]&=&(m-n)L_{m+n}+{D\over 12}(m^3 -m)\Kr{m+n}~,\\
\L[L_m,I_n\R]&=&-nI_{m+n}-\half Q_s (m^2+m)\Kr{m+n}~,\\
\L[I_m,I_n\R]&=&-m\Kr{m+n}~.\EA\label{hat}\EE

\noi
where

  \BE  L_m = L_m^{matter}-\half\sum_n:I_{m-n}I_n: +\ \half Q_s(m+1)I_m\EE

\noi
and  $D=26 \  (D=2) ,\   Q_s=\sqrt{{25-d\over3}}
 \  (Q_s=\sqrt{{1-d\over3}})$
for the DDK { }(KM)  realization. The states $\ket\Ups$ above
are thus primaries of the DDK and KM CFT's respectively.

\vskip .2in

We are now ready  to establish four theorems concerning the DDK and KM
realizations of topological descendants (two of them concern
null vectors exclusively).

Let us take a topological descendant and split its topological generators into
their DDK or KM components.  The result is expected to be a sum of ghost-free
terms,  matter and/or scalar-ghost mixed terms, and pure-ghost terms.
This is indeed what happens in the general  case.  However, there are
special cases in which {\it all the ghosts cancel out}, so that the DDK or KM
realization reduces  the topological descendant to a secondary
state of the DDK or KM  CFT. Those special cases are considered
in the following theorems.

\vskip .2in

Theorem 1.-  Let $\ket\Xi$ be a $\cG_0$-invariant topological
descendant annihilated
by all the positive modes $\cG_{n>0}$, and with ghost number equal
to the one
assigned to $c_1 \ket0_{gh}$ (whatever the convention). Then the DDK
realization reduces $\ket\Xi$  to

 \BE  \ket\Xi_{DDK} = \ket{\Psi_{DDK}} \otimes c_1 \ket0_{gh}  \label{XDDK} \EE

\noi
where $\ket{\Psi_{DDK}}$ is a descendant of the DDK CFT.

\vskip .2in

Proof.-    Let $\ket\Xi$ be annihilated by $\cG_{n \geq 0}$.
  In the DDK realization this
results in  $b_{n \geq 0}\   \ket\Xi_{DDK} = 0$.
  Thus $\ket\Xi_{DDK}$ cannot contain
any $c_{n \leq 0} $ ghost modes, since $\{b_m, c_n\} = \delta_{m+n}$.
 Nor can it
contain any annihilation modes $c_{n>1}$ or $b_{n>-2}$ . Therefore in the
ghost sector only $c_1$ and $b_{n<-1}$ modes are possible.  But
$\ket\Xi$ is  by assumption in the same ghost-number subspace
 as $c_1 \ket0_{gh}$,
so that  $\ket\Xi_{DDK}$  must have the form \req{XDDK}.

\vskip .2in

Theorem 2.- Let  $\ket\Xi$ be a $\cQ_0$-invariant topological
descendant annihilated
by all the positive modes $\cQ_{n>0}$, and with ghost number equal to the one
assigned to $\ket0_{gh}$. Then the KM realization reduces $\ket\Xi$ to

     \BE  \ket\Xi_{KM} = \ket{\Psi_{KM}} \otimes \ket0_{gh}  \label{XKM} \EE

\noi
where $\ket{\Psi_{KM}}$ is a descendant of the KM CFT.

\vskip .2in

Proof.- Let $\ket\Xi$ be annihilated by $\cQ_{n \geq 0}$.
 In the KM realization this
results in $b_{n \geq 0}\  \ket\Xi_{KM} = 0$.
 Thus $\ket\Xi_{KM}$ cannot contain
any $c_{n \leq 0}$ modes, nor can it contain
  any annihilation modes $c_{n > 0}$ or $b_{n>-1}$ either.
Thus, only $b_{n<0}$ modes are allowed
 in the ghost sector. But  $\ket\Xi$ is
by assumption in the same ghost-number subspace as $\ket0_{gh}$, so that
$\ket\Xi_{KM}$ must have the form \req{XKM} .

\vskip .2in

For the particular case of null vectors,
 it is now rather easy to prove the following.

\vskip .2in

Theorem 3.- Let $\ket\Xi$ be a $\cG_0$-invariant
 topological null vector at level $l$ with ghost
number equal to the one assigned to $c_1 \ket0_{gh}$ .
 Then the DDK realization reduces
$\ket\Xi$ to a level $l$ null vector of the DDK CFT.

\vskip .2in

Proof.-  Applying Theorem 1 $\ \ket\Xi$  must be of the form \req{XDDK}.
Then the DDK realization translates straightforwardly
 the topological highest weight conditions
  $\cL_{n>0} \ket\Xi = \cH_{n>0} \ket\Xi = \cQ_{n>0} \ket\Xi =
\cG_{n>0} \ket\Xi = 0 $
 into the highest weight conditions of the DDK CFT
  $L_{n>0} \ket{\Psi_{DDK}} =  I_{n>0} \ket{\Psi_{DDK}} = 0$.
In addition, from

\BE (\cL_0 \ket\Xi)_{DDK} =  L_0 \ket{\Psi_{DDK}} \otimes c_1 \ket0_{gh} +
\ket{\Psi_{DDK}} \otimes l_0 \ c_1 \ket0_{gh} \EE

\noi
we obtain  $\Delta_{\ket\Psi} = l + 1 $.
 Since the conformal weight of all the
primaries in the DDK CFT is equal to $1$
 (DDK dressing), we conclude that
$\ket{\Psi_{DDK}}$ is a null vector at level $l$.

\vskip .2in

Theorem 4.- Let $\ket\Xi$ be a $\cQ_0$-invariant
 topological null vector at level $l$ with ghost
number equal to the one assigned to $\ket0_{gh}$.
 Then the KM realization reduces
$\ket\Xi$ to a level $l$ null vector of the KM CFT.

\vskip .2in

Proof.- Applying Theorem 2 $\ \ket\Xi$ must be of
 the form \req{XKM}.  The KM
realization translates straightforwardly the topological
 highest weight conditions on $\ket\Xi$ into the
highest weight conditions of the KM CFT
 on $\ket{\Psi_{KM}}$. Moreover, since

\BE (\cL_0 \ket\Xi)_{KM} = L_0 \ket{\Psi_{KM}} \otimes \ket0_{gh} \EE

\noi
and the conformal weight of all the primaries
 in the KM CFT is zero  (KM dressing),
we conclude that $\ket{\Psi_{KM}} $ is a null vector at level $l$.

\section{Level $2$ Topological Descendants}\lvm

Level $2$ topological descendants of bosonic type have the generic form

\BE \ket\Xi^B = (\a \cL_{-1}^2 + \th \cL_{-2} +
 \G \cH_{-1} \cL_{-1} + \b \cH_{-1}^2 +
 \g \cH_{-2} + \delta \cQ_{-1} \cG_{-1} )\ \ket\P_\htop  \label{bosdes} \EE

\noi
where $\ket\P_\htop$ is a chiral primary state
 with $\cH_0$ eigenvalue $\htop$.

{}From now on we will focus on the special cases of
 $\cG_{n \geq0}$ -invariant  states
and  $\cQ_{n \geq 0}$ - invariant  states.

\vskip .2in

Let us start with the $\cG_{n \geq 0}$ - invariant
 descendants $\ket\Xi^{BG}$. The condition
$\cG_0 \ket\Xi^{BG} = 0 $ results in the equations

\BE  2 \a - \G + 2 \delta = 0,\ \ \G + 2 \delta - 2 \b = 0, \ \
2 \th + \G - \b -\g + 2 \delta  = 0  \label{G0inv} \EE

\noi
while $\cG_1 \ket\Xi^{BG} = 0$ gives

\BE  2 \a + 3 \th - \G + \b -\g +
 2 \delta \ (2 + \htop + {\ctop\over3}) = 0\ . \label{G1inv} \EE

\noi
The conditions $\cG_{n>1} \ket\Xi^B= 0$ are satisfied identically (on
any $\ket\Xi^B$ ).

\vskip .2in

Now let us consider the $\cQ_{n \geq 0} $ - invariant
bosonic descendants $\ket\Xi^{BQ}$. The condition
$\cQ_0 \ket\Xi^{BQ} = 0$ gives

\BE  \b = 0, \quad \g = 0, \quad \G = 2\delta  \label{Q0inv} \EE

\noi
and  $\cQ_1 \ket\Xi^{BQ} = 0$ results in

\BE  \th = -2 \htop \delta \ . \label{Q1inv} \EE

 The conditions $\cQ_{n>1} \ket\Xi^B = 0$
are satisfied identically.

\vskip .2in

As a first application of these results, notice that
$\cQ_0$ - invariance (BRST- invariance) plus
$\cG_0$ - invariance imply the vanishing of all the
coefficients in $\ket\Xi^B$. That is, a $\cQ_0$ and $\cG_0$ - invariant
state must be a chiral primary necessarily (as we already knew).

\vskip .2in

Let us now move onto the fermionic descendants.
 Since $\cG_n$ and $\cQ_n$ have opposite $\cH_0$ charges,
there are two different types of fermionic descendants,
mirrored under $\cG_n$ versus $\cQ_n$.
 The fermionic secondary state with
$\cH_0$ eigenvalue $\htop+1$ has the form

\BE \ket\Xi^{FG} = (\delta \cL_{-1} \cG_{-1} + \cG_{-2} +
  \b \cH_{-1} \cG_{-1}) \ket\P_\htop \label{ferGdes} \EE

\noi
and satisfies the conditions $\cG_{n \geq0} \ket\Xi^{FG} = 0$
identically, in particular $\cG_0$ - invariance.

The fermionic secondary state with $\cH_0$ eigenvalue
 $\htop - 1$ has the "mirrored" form

\BE \ket\Xi^{FQ} = (\a \cL_{-1} \cQ_{-1} + \cQ_{-2} +
  \g \cH_{-1} \cQ_{-1}) \ket\P_\htop \label{ferQdes} \EE

\noi
and satisfies the conditions $\cQ_{n \geq0} \ket\Xi^{FQ} = 0$
identically, in particular $\cQ_0$ - invariance.

Notice that fermionic descendants, at level 2, are either of the
$\ket\Xi^{FG}$ type or rather of the $\ket\Xi^{FQ}$ type; in other
words, not only there are not $Q_0$ plus $G_0$ - invariant
 fermionic descendants, like in the bosonic case,
but not even fermionic descendants of the type $\ket\Xi^F$
(neither $\cG_{n \geq0}$ nor $\cQ_{n \geq0}$ -
invariant).

\vskip .2in

When the descendants are null states, one has to impose the
 complete set of highest weight conditions as well. We will present
an analysis of  level 2 and level 3 topological null states
in \cite{BJIS2}.

\subsection{DDK  Realization of the Level $2$ Topological Descendants}\lvm

The general analysis of the DDK and KM realizations of
topological descendants, with the resulting ghosts
structures, etc...., is beyond the scope of this letter.
Here we will consider only the cases met by Theorem 1
(for DDK) and Theorem 2 (for KM).

Let us start with the DDK realization of
bosonic descendants. As we mentioned before, in the DDK
realization the secondary states are built on chiral
primaries of the form $\ket\P_\htop = \ket\Ups \otimes
c_1 \ket0_{gh}$ , where $\ket\Ups$ is a primary state
of the DDK CFT. Then the DDK realization of any given $\ket\Xi^B$
results in the following terms:

a) Terms with the structure $[{\rm ghost-free}]\ket\Ups
 \otimes c_1 \ket0_{gh}$,
 where $[{\rm ghost-free}]$ is

\BE \a L^2_{-1}+(\th+2\delta)L_{-2}-\G \sqrt{3-\ctop\over3}\ I_{-1}L_{-1}+
\b\ {3-\ctop\over3}\ I^2_{-1}-(\g+2\delta)\sqrt{3-\ctop\over3}\ I_{-2}\ .\EE

b) Terms with the structure $[{\rm ghost-free}]\ket\Ups \otimes c_0
\ket0_{gh}$,
 where $[{\rm ghost-free}]$ is

\BE   (2\a - \G + 2\delta) L_{-1} -
    (\G + 2\delta - 2\b) \sqrt{3-\ctop\over3}\   I_{-1}  \ . \EE

c) Pure ghost terms with the structure $\ket\Ups \otimes
 [{\rm ghost}] \ket0_{gh}$, where $[{\rm ghost}]$ is

\BE  (-2\th-\G+\b+\g-2\delta)\  b_{-2}c_0c_1  +
     (2\a+3\th-\G+\b-\g+2\delta\ (2+\htop+{\ctop\over3}))\  c_{-1} . \EE

Inspecting eqn's \req{G0inv} - \req{G1inv} we see that by imposing $\cG_0$
and $\cG_1$ invariance all the coefficients in b) and c) vanish.
Therefore, the structure of the $\cG_{n \geq0}$ - invariant
bosonic descendants in the DDK realization is

\BE \ket\Xi^{BG}_{DDK} = \ket{\Psi_{DDK}} \otimes c_1 \ket0_{gh}  \EE

\noi
where $\ket{\Psi_{DDK}}$ is a secondary state of the DDK CFT,
 in agreement with Theorem 1.

For the $\cG_{n \geq0}$ - invariant fermionic descendants,
it is not possible to meet the conditions of Theorem 1, so
it is not of much use in this case. However, using relation
\req{QG} it is indeed possible to investigate the structure of the
fermionic states directly connected, through $\cQ_0$,
to the given $\cG_{n \geq0}$ - invariant bosonic states. These fermionic
descendants are of the $\cQ_{n \geq0}$ - invariant type.
In the case at hand, relation \req{QG} reads

\BE \ket\Xi^{BG} = {1\over4} \cG_0 \cQ_0 \ket\Xi^{BG} =
    {1\over4} \cG_0 \ket\Xi^{FQ} \ .\EE

\noi
In the DDK realization this relation becomes

\BE \ket\Xi^{BG}_{DDK} = {1\over4} b_0 \ket\Xi^{FQ}_{DDK} \EE

\noi
that shows that
$  \ket\Xi^{FQ}_{DDK} = 4 c_0 \ket\Xi^{BG}_{DDK} \ $.
Therefore, in the DDK realization the $\cQ_{n\geq0}$ - invariant fermionic
descendants reduce to the structure

\BE \ket\Xi^{FQ}_{DDK} = \ket{\Psi_{DDK}}
 \otimes c_0 c_1 \ket0_{gh} \ . \label{FQDDK}\EE

\subsection{KM  Realization of the Level $2$ Topological Descendants}\lvm

In the KM realization the secondary states are built on chiral
primaries of the form $\ket\P_\htop = \ket\Ups \otimes \ket0_{gh}$ ,
where $\ket\Ups$ is a primary state of the KM CFT. As a result, the
KM realization of a bosonic descendant $\ket\Xi^B$ is given by
the following terms:

a) Terms with the structure $[{\rm ghost-free}]\ket\Ups \otimes \ket0_{gh}$,
 where $[{\rm ghost-free}]$ is

\BE \a L^2_{-1}+\th L_{-2}+\G \sqrt{3-\ctop\over3}\ I_{-1}L_{-1}+
  \b\ {3-\ctop\over3}\ I^2_{-1}+\g\sqrt{3-\ctop\over3}\ I_{-2}\ . \EE

b) Terms with the structure
 $[{\rm ghost-free}]\ket\Ups \otimes b_{-1} c_0 \ket0_{gh}$,
 where $[{\rm ghost-free}]$ is

\BE   (- \G + 2\delta) L_{-1}  -  2\b  \sqrt{3-\ctop\over3}\  I_{-1}
    \ . \EE

c) Pure ghost terms with the structure $\ket\Ups \otimes
 [{\rm ghost}] \ket0_{gh}$, where $[{\rm ghost}]$ is

\BE  (\b-\g)\  b_{-2}c_0  -
     (\th+\b+\g+2\delta\htop)\ b_{-1}c_{-1} . \EE

Imposing $\cQ_0$ and $\cQ_1$ invariance, eqn's \req{Q0inv} -
\req{Q1inv}, all the coefficients in b) and c) vanish.
Therefore, the structure of the $\cQ_{n\geq0}$ - invariant
bosonic descendants in the KM realization is

\BE \ket\Xi^{BQ}_{KM} = \ket{\Psi_{KM}} \otimes \ket0_{gh} \EE

\noi
where $\ket{\Psi_{KM}}$ is a secondary state of the KM CFT,
in agreement with Theorem 2.

For the $\cQ_{n \geq0}$ - invariant fermionic descendants it is
not possible to satisfy the conditions of Theorem 2. However,
we saw in the previous subsection that the DDK realization of
 those states actually reduces to the form \req{FQDDK}.

The same reasoning applies now to the $\cG_{n \geq0}$ - invariant
fermionic descendants, connected through $\cG_0$ to the given
$\cQ_{n \geq0}$ - invariant bosonic states. In this case,
relation \req{QG} reads

\BE \ket\Xi^{BQ} = {1\over4} \cQ_0 \cG_0 \ket\Xi^{BQ} =
    {1\over4} \cQ_0 \ket\Xi^{FG} \EE

\noi
so that

\BE \ket\Xi^{BQ}_{KM} = {1\over4} b_0 \ket\Xi^{FG}_{KM} \ . \EE

Therefore
$  \ket\Xi^{FG}_{KM} = 4 c_0 \ket\Xi^{BQ}_{KM} \ $. As a result,
in the KM realization the $\cG_{n \geq0}$ - invariant fermionic
descendants reduce to the structure

\BE \ket\Xi^{FG}_{KM} = \ket{\Psi_{KM}} \otimes c_0 \ket0_{gh} \ . \EE

\section{Final Remarks}\lvm

We have analyzed the DDK and KM realizations of topological
secondary states. In particular we have investigated the issue
of the reduction of topological descendants to simple structures
(ghost-free or quasi ghost-free structures). To this purpose
four {\it no-ghost} theorems have been derived, two of them
for null states. Finally, we have analyzed the corresponding
 results for the case of level 2 topological descendants,
showing complete agreement with the {\it no-ghost} theorems.

The relation between the topological states considered in this
letter and Lian-Zuckerman states \cite{LZ} is under investigation
\cite{BJIS3}.

The particular case of topological null descendants will be
considered in \cite{BJIS2}.

\vskip 1cm

\centerline{\bf Acknowledgements}

We would like to thank Aliosha Semikhatov and Bert Schellekens
 for many discussions and for carefully reading the manuscript.

\end{document}